\documentclass[twocolumn,showpacs,preprintnumbers,amsmath,amssymb,nofootinbib]{revtex4}

\usepackage{graphicx}
\usepackage{dcolumn}
\usepackage{bm}

\begin{document}

\preprint{APS/123-QED}

\title{Effects of hyperonic many-body force on $B_\Lambda$ values of hypernuclei} 

\author{M.\ Isaka$^{1}$}
\author{Y.\ Yamamoto$^{2}$}
\author{Th.A.\ Rijken$^{3}$$^{2}$}
\affiliation{
$^{1}$Research Center for Nuclear Physics (RCNP), Osaka University, Ibaraki, Osaka, 567-0047, Japan\\
$^{2}$Nishina Center for Accelerator-Based Science,
Institute for Physical and Chemical
Research (RIKEN), Wako, Saitama, 351-0198, Japan\\
$^{3}$IMAPP, Radboud University, 6500 GL Nijmegen, The Netherlands
}

\date{\today}

\begin{abstract}
The stiff equation of state (EoS) giving the neutron-star mass of $2M_{\odot}$ suggests 
the existence of strongly repulsive many-body effect (MBE) not only in nucleon channels 
but also in hyperonic ones.
As a specific model for MBE, the repulsive multi-pomeron exchange potential (MPP)
is added to the two-body interaction together with the phenomenological three-body 
attraction. For various versions of the Nijmegen interaction models, the MBE parts 
are determined so as to reproduce the observed data of $B_\Lambda$.
The mass dependence of $B_\Lambda$ values is shown to be reproduced well by adding MBE 
with the strong MPP repulsion assuring the stiff EoS of hyperon-mixed neutron-star matter, 
when $P$-state components of the adopted interaction model lead to 
almost vanishing contributions.
The nuclear matter $\Lambda\!N$ $G$-matrix interactions are derived and used in $\Lambda$
hypernuclei on the basis of the averaged-density approximation (ADA).
The $B_\Lambda$ values of hypernuclei with $9 \le A \le 59$ are analyzed
in the framework of Antisymmetrized Molecular Dynamics with use of the two types of 
$\Lambda\!N$ $G$-matrix interactions including strong and weak MPP repulsions. 
The calculated values of $B_\Lambda$ reproduce the experimental data finely within 
a few hundred keV. The values of $B_\Lambda$ in $p$-states also can be reproduced well, 
when ADA is modified to be suitable also to weakly-bound $\Lambda$ states. 
\end{abstract}

\pacs{Valid PACS appear here}
\maketitle

\section{Introduction}

$\Lambda$ binding energies $B_\Lambda$ are basic quantities
in $\Lambda$ hypernuclei.
In 1950's the values of $B_\Lambda$ were extracted from 
$\Lambda$ hypernuclei with mass $A<16$ observed in emulsion.
After 1980's, medium and heavy $\Lambda$ hypernuclei have been 
produced by counter experiments such as $(\pi^+,K^+)$ reactions.
Recently accurate data of $B_\Lambda$ values in ground and excited states 
of hypernuclei have been obtained by $\gamma$-ray observations and
$(e,e'K^+)$ reactions. 
On the other hand, theoretical baryon-baryon interaction models have 
been developed~\cite{NSC89,NSC97,ESC04,ESC08,Hol89,Reu94,Fuji07}, 
where $\Lambda\!N$-$\Sigma\!N$ coupling terms have been included in order to
reproduce values of $\Lambda$ single particle potentials $U_\Lambda$ values 
in nuclear matter more or less realistically.
Because hyperon($Y$)-nucleon($N$) scattering data are extremely limited,
there remain remarkable ambiguities in $Y\!N$ interaction models:
Values of $U_\Lambda$ for various interaction models are substantially
different from each other.

The $Y\!N$ interactions are related intimately to the recent topic
in neutron-stars. The large observed masses of 
$2M_{\odot}$~\cite{Demorest10,Antoniadis13} give a severe condition 
for the stiffness of equation of state (EoS) of neutron-star matter.
The stiff EoS giving the maximum mass of $2M_{\odot}$ can be derived 
from the existence of strong three-nucleon repulsion (TNR) in the 
high-density region. However, the hyperon ($Y$) mixing in neutron-star 
matter brings about the remarkable softening of the EoS, which 
cancels the TNR effect for the maximum mass~\cite{Baldo00,Vidana00,NYT}.
This problem is known as the ``Hyperon puzzle". It is considered that 
this puzzle can be solved if strong repulsions exist not only in $N\!N\!N$ 
cannels but also in $Y\!N\!N$ and $Y\!Y\!N$ channels~\cite{NYT}.

Recently, there have been reported the trials to extract the $\Lambda\!N\!N$
repulsions from the systematic data of $B_\Lambda$~\cite{YFYR14,Lonard14}.
In Refs.\cite{YFYR13,YFYR14}, the multi-pomeron exchange potential (MPP) 
was added to the two-body baryon-baryon interaction $V_{BB}$ 
together with the phenomenological three-body attraction (TBA).
Then, the parameters included in MPP and TBA were determined so as to reproduce
the angular distribution of $^{16}$O+$^{16}$O scattering at $E/A=70$ MeV and 
the nuclear saturation property, where the MPP contributions were decisive to 
reproduce the experimental angular distribution and brought about the stiff
EoS enough to give maximum masses over $2M_{\odot}$.
$V_{BB}$ gives the potentials in $NN$ and $Y\!N$ channels, and MPP is universal
in all baryon channels. The TBA parts in $Y\!N$ channels are determined
so as to reproduce hypernuclear data reasonably.
On the basis of this ($V_{BB}$+MPP+TBA) model, it was shown that
the EoS was still stiff enough to reproduce neutron stars with $2M_\odot$
in spite of substantial softening by hyperon mixing.

The aim of this work is to investigate the $\Lambda\!N$ sectors of
the $V_{BB}$+MPP+TBA model, especially the many-body effects (MBE) given 
by MPP+TBA parts, through structure calculations of $\Lambda$ hypernuclei
within the framework of the antisymmetrized molecular dynamics for hypernuclei (HyperAMD).
When we determine the MPP+TBA part so as to reproduce the experimental
values of $B_\Lambda$, it is evident that this part is dependent on
the interaction model for $V_{BB}$. In other words, it is indispensable that
reliable interaction models should be used for $V_{BB}$ in order to investigate MBE.
We start from the Nijmegen interaction models for $V_{BB}$, being reliable enough to 
extract MBE in spite of remained ambiguity for reproducing values of $B_\Lambda$.

In Refs.\cite{Lonard14,Lonard15}, the strengths of $\Lambda\!N\!N$ forces 
were determined by the fitting procedure to the data of $B_\Lambda$.
Their $\Lambda\!N\!N$ repulsion in the best fitting case seems to be
abnormally strong. The reason seems to be because they start from the 
two-body $\Lambda\!N$ interaction with no $\Lambda\!N$-$\Sigma\!N$ term, 
giving an overbinding value of $U_\Lambda$. 

In our case, the $\Lambda\!N$-$\Sigma\!N$ coupling terms are included 
in the Nijmegen models so that their strengths are determined to 
reproduce physical observables through channel-coupling effects to 
$\Lambda\!N$-$\Lambda\!N$ diagonal channels.  
Then, there remains a rather small room for MBE around normal-density region, 
where the MPP and TBA contributions are cancelled substantially with each other.
In our previous work~ \cite{IYR16}, referred to I, the experimental values of 
$B_\Lambda$ have been reproduced systematically by the HyperAMD calculations using
a special Nijmegen model having only a very small room for MBE.
In Ref.~\cite{IYR16}, even in this case it was demonstrated that the small MBE works
to improve the fitting of $B_\Lambda$ values to experimental data.
In this work, we show that they appear more clearly in the case of using  
the updated versions of Nijmegen extended-soft core (ESC) models.  

This paper is organized as follows. 
In the next section, the various versions of the Nijmegen models and MBE (MPP+TBA) 
are explained, and the $\Lambda\!N$ $G$-matrix calculations are performed.
Different features of interaction models are discussed by showing $U_\Lambda$
values in nuclear matter.
In Sec. III, the detailed analysis for $\Lambda$ hypernuclei with $9 \le A \le 59$
are performed on the basis of HyperAMD with use of $G$-matrix interactions including MBE.
It is discussed what feature of the two-body interaction allows the existence of 
strong repulsion suggested by the stiff EoS of neutron stars.
Section V summarizes this paper.

\section{$U_\Lambda$ in nuclear matter}
\label{SecII}

\subsection{Nijmegen interaction models}

The meson-theoretical models for $Y\!N$ interactions
have been developed continuously by the Nijmegen group.
In the earlier stage, they developed the hard-core models~\cite{NDF}
(NHC-D and -F) and the soft-core model (NSC89)~\cite{NSC89}.
After that, the trial started to take into account the $G$-matrix results
in the modeling of $Y\!N$ interactions.
As the first outcome of this approach, the NSC97 models~\cite{NSC97}
were proposed, where the six versions a$\sim$f were designed so as
to be of different strengths of the $\Lambda\!N$ spin-spin parts.
Then, the observed splitting energies of spin-doublet states
in $\Lambda$ hypernuclei suggested that the spin-spin strengths 
of NSC97e and NSC07f were in a reasonable region.
Epoch-making development of the Nijmegen models
was accomplished by the ESC models,
in which two-meson and meson-pair exchanges are taken into account
explicitly. In the one-boson exchange (OBE) models these effects are 
implicitly and roughly described by exchanges of `effective mesons'.
After some trial versions, there appeared the specific versions 
ESC04a/b/c/d~\cite{ESC04}, features of which were very different 
from those of the OBE models especially in $S=-2$ channels.
However, there remain some serious problems in NSC97 and ESC04 models.
The first is that the derived values of $\Lambda$ spin-orbit
splitting energies are too large in comparison with
the experimental values. The second is that the derived
$\Sigma$-nucleus potentials $U_\Sigma$ are attractive, whereas
the experimental values are indicated to be repulsive.
Furthermore, the $\Xi\!N$ interactions seem to be unreliable:
The $U_\Xi$ values derived from the NSC97 (ESC04a/b) models
are strongly (weakly) repulsive. Those for ESC04c/d are 
attractive, but their partial-wave contributions 
seem to be rather problematic.
These problems have been further investigated in ESC08a/b/c~\cite{ESC08} where the treatments for axial-vector and pair terms are improved, and the effects of the quark Pauli-forbidden states in the repulsive-core representation are taken into account. However, in these models the $\Xi N$ cross sections are too large. At present the possibility is investigated to replace in $U_{\Xi}$ part of the two-body attraction by a three-body force contribution. 
This trial to improve the $\Xi N$ part does not affect the $NN$ and $YN(S=-1)$ parts.

Here, in order to investigate sizes of MBE needed for different interaction models,
we pick up ESC08a/b/c, ESC04a and NSC97e/f among the various the Nijmgen models.
In Ref.~\cite{IYR16}, we used ESC08c in an early stage of parameter fitting:
This version denoted as ESC08c(2012)~\cite{ESC2012}, and the recent version 
as ESC08c(2014)~\cite{ESC08c1,ESC08c2,ESC08c3}. Hereafter, ESC08c(2012) and 
ESC08c(2014) are denoted simply as ESC12 and ESC14, respectively.
These two versions of ESC08c are used in this work
mainly.
Very recently, there is given the latest version ESC08c(2016) \cite{ESC16}, ESC16,
though it is not used in this work.
%
The reason why we pick up ESC04a (NSC97e/f) among ESC04a/b/c/d (NSC97a/b/c/d/f)
is because ESC04b/c/d give more attractive values of $U_\Lambda$ than ESC04a, and
NSC97a/b/c are with unreasonable spin-spin parts, not giving binding
of $^3_\Lambda$H. 

One of the ideas to avoid remarkable softening of neutron-star EoS by hyperon 
mixing is to assume that the strong three-body repulsions work universally 
for $Y\!N\!N$, $Y\!Y\!N$ $Y\!Y\!Y$ as well as for $N\!N\!N$ \cite{NYT}.
As a model of universal repulsions among three and four baryons,
we introduce the multi-pomeron exchange potential (MPP).
Additionally to MPP, the three-baryon attraction (TBA) is assumed
phenomenologically. MPP and TBA in nucleon channels are determined
so as to reproduce the experimental angular distributions of 
$^{16}$O+$^{16}$O elastic scattering ($E/A$=70 MeV) and the nuclear
saturation property. In hyperonic channels, they should be taken
consistently with hypernuclear data:
For each interaction model $V_{BB}$, MPP and TBA parts are adjusted 
so as to reproduce experimental data of $B_\Lambda$ as well as possible.

The specific form of MPP is given as the $N$-body local potential by 
pomeron exchange $W^{(N)}({\bf x}_1, ..., {\bf x}_N)$ \cite{YFYR13,YFYR14}
and the effective two-body potential in a baryonic medium is obtained
by integrating over the coordinates ${\bf x}_3,..., {\bf x}_N$; 
\begin{eqnarray}
\nonumber
&& V_{MPP}^{(N)}({\bf x}_1,{\bf x}_2) \\
&&  = \rho_{}^{N-2} 
  \int\!\! d^3\!x_3 ... \int\!\! d^3\!x_N\ 
  W^{(N)}({\bf x}_1,{\bf x}_2, ..., {\bf x}_N)
  \nonumber \\ 
&&  =g_P^{(N)} g_P^N\frac{\rho^{N-2}}{{\cal M}^{3N-4}}\cdot
  \left(\frac{m_P}{\sqrt{2\pi}}\right)^3
  \exp\left(-\frac{1}{2}m_P^2 r_{12}^2\right).
 \label{eq:2}
 \end{eqnarray}
We assume that
the dominant mechanism is triple and quartic pomeron exchange.
The values of the two-body pomeron strength $g_P$ and 
the pomeron mass $m_P$ are taken from the adopted ESC model.
A scale mass ${\cal M}$ is taken as a proton mass.
TBA is assumed as a density-dependent two-body interaction
\begin{eqnarray}
\nonumber
&& V_{TBA}(r;\rho)\\
&& = V_{0}\, \exp(-(r/2.0)^2)\, \rho\, 
\exp(-\eta \rho)\, (1+P_r)/2 \ ,
\label{eq:4}
\end{eqnarray}
$P_r$ being a space-exchange operator.
There are given the three sets with different strengths of 
MPP \cite{YFYR13,YFYR14}. We consider the set MPa as a guidance in this work,
where the parameters are taken as
$g_P^{(3)}=2.34$, $g_P^{(4)}=30.0$, $V_0=-32.8$ and $\eta=3.5$. 
Then, the most important is whether or not such a strongly repulsion
given by these values of $g_P^{(3)}$ and $g_P^{(4)}$ is allowable in
reproducing the mass dependence of $B_\Lambda$ values.

\subsection{$G$-matrix interaction}
\label{Sec:Gmat}

We start from the channel-coupled $G$-matrix equation 
for the baryon pair $B_1 B_2$ in nuclear matter~\cite{Yam10}, 
where $B_1B_2 = \Lambda\!N$ and $\Sigma\!N$: 
\begin{eqnarray}
G_{cc_0}=v_{cc_0} + \sum_{c'} 
 v_{cc'} {Q_{y'} \over \omega -\epsilon_{B'_1}-\epsilon_{B'_2} +\Delta_{yy'}}
G_{c' c_0} \ ,
\label{eq:GM1}
\end{eqnarray}
where $c$ denotes a $Y\!N$ relative state $(y, T, L, S, J)$ with $y=(B_1B_2)$. 
$S$ and $T$ are spin and isospin quantum numbers, respectively.
Orbital and total angular momenta are denoted by $L$ and $J$,
respectively, with ${\bf J}={\bf L}+{\bf S}$.
Then, a two-particle state is represented as $^{2S+1}L_J$.
In Eq.~(\ref{eq:GM1}), $\omega$ gives the starting energy in 
the channel $c_0$.
$\Delta_{yy'}= M_{B_1}+M_{B_2}-M_{B'_1}-M_{B'_2}$ denotes 
the mass difference between two baryon channels.
The Pauli operator $Q_y$ acts on intermediate nucleon states
in a channel $y=(B_1B_2)=(\Lambda\!N$ and $\Sigma\!N$).
The continuous (CON) choice is adopted for intermediate 
single particle potentials in the $G$-matrix equation.

The $G$-matrix equation~(\ref{eq:GM1}) is represented in the
coordinate space, whose solutions give rise to $G$-matrix elements.
The hyperon single particle (s.p.) energy $\epsilon_Y$ 
in nuclear matter is given by
\begin{eqnarray}
\epsilon_Y(k_Y)={\hbar^2k_Y^2 \over 2M_Y} + U_Y(k_Y) \ ,
\label{eq:GM2}
\end{eqnarray}
where $k_Y$ is the hyperon momentum.
The potential energy $U_Y$ is obtained self-consistently
in terms of the $G$-matrix as
\begin{eqnarray}
&& U_Y(k_Y) = 
\nonumber
\\
&& \sum_{|{\bf k}_N|} \langle {\bf k}_Y {\bf k}_N
\mid G_{YN}(\omega=\epsilon_Y+\epsilon_N) \mid
{\bf k}_Y {\bf k}_N \rangle 
\label{eq:GM3}
\end{eqnarray}

In Table~\ref{Gmat-1}, we show the potential energies $U_\Lambda$
of a zero-momentum $\Lambda$ at normal density $\rho_0$ 
($k_F$=1.35 fm$^{-1}$). Their $S$- and $P$-state contributions
are given by $U_\Lambda(S)$ and $U_\Lambda(P)$, respectively.
They are calculated for adopted Nijmegen models.
It is noted that the $U_\Lambda$ values for ESC08a/b, ESC14 and ESC12 are 
rather similar to each other, and those for NSC97e/f (ESC04a) are less (more)
attractive due to strongly repulsive (attractive) $P$-state contributions.
As is given in \cite{ESC16}, we have $U_\Lambda=-39.6$ MeV, 
$U_\Lambda(S)=-38.8$ MeV and $U_\Lambda(P)=+0.7$ MeV for ESC16.
It's $S$-($P$-) contribution is slightly less attaractive (more repulsive)
than those for ESC14.

\begin{table}
\caption{$\Lambda$ potential energies $U_\Lambda$ [MeV] at normal density
for adopted interaction models. $U_\Lambda(S)$ and $U_\Lambda(P)$  
are $S$- and $P$-state contributions, respectively, in unit of MeV.}
  \label{Gmat-1}
  \begin{ruledtabular}
  \begin{tabular}{lccc}
     & $U_\Lambda$  &  $U_\Lambda(S)$ &  $U_\Lambda(P)$  \\
  \hline
  ESC08a  & $-40.6$ & $-39.5$  & $+0.5$ \\
  ESC08b  & $-39.4$ & $-37.0$  & $-0.6$ \\
  ESC14   & $-40.8$ & $-39.6$  & $+0.4$ \\
  ESC12   & $-40.0$ & $-40.0$  & $+1.5$ \\
  ESC04a  & $-43.2$ & $-38.4$  & $-3.7$ \\
  NSC97e  & $-37.7$ & $-40.4$  & $+4.0$ \\
  NSC97f  & $-34.8$ & $-39.1$  & $+5.6$ \\
  \end{tabular}
  \end{ruledtabular}
\end{table}

\begin{figure}[htb] 
\centering 
\includegraphics[width=8cm]{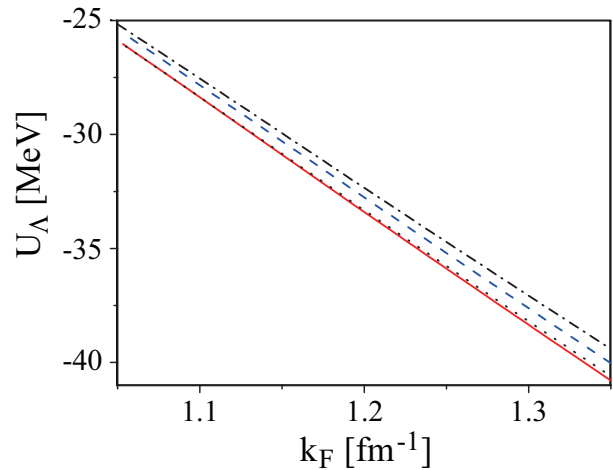} 
\caption{\small \label{UL}
(Color online) $U_\Lambda$ as a function of $k_F$.
Solid, dashed, dotted and dot-dashed curves are for
ESC14, ESC12, ESC08a and ESC08b, respectively.
}
\end{figure} 

\begin{figure}[htb] 
\centering 
\includegraphics[width=8cm]{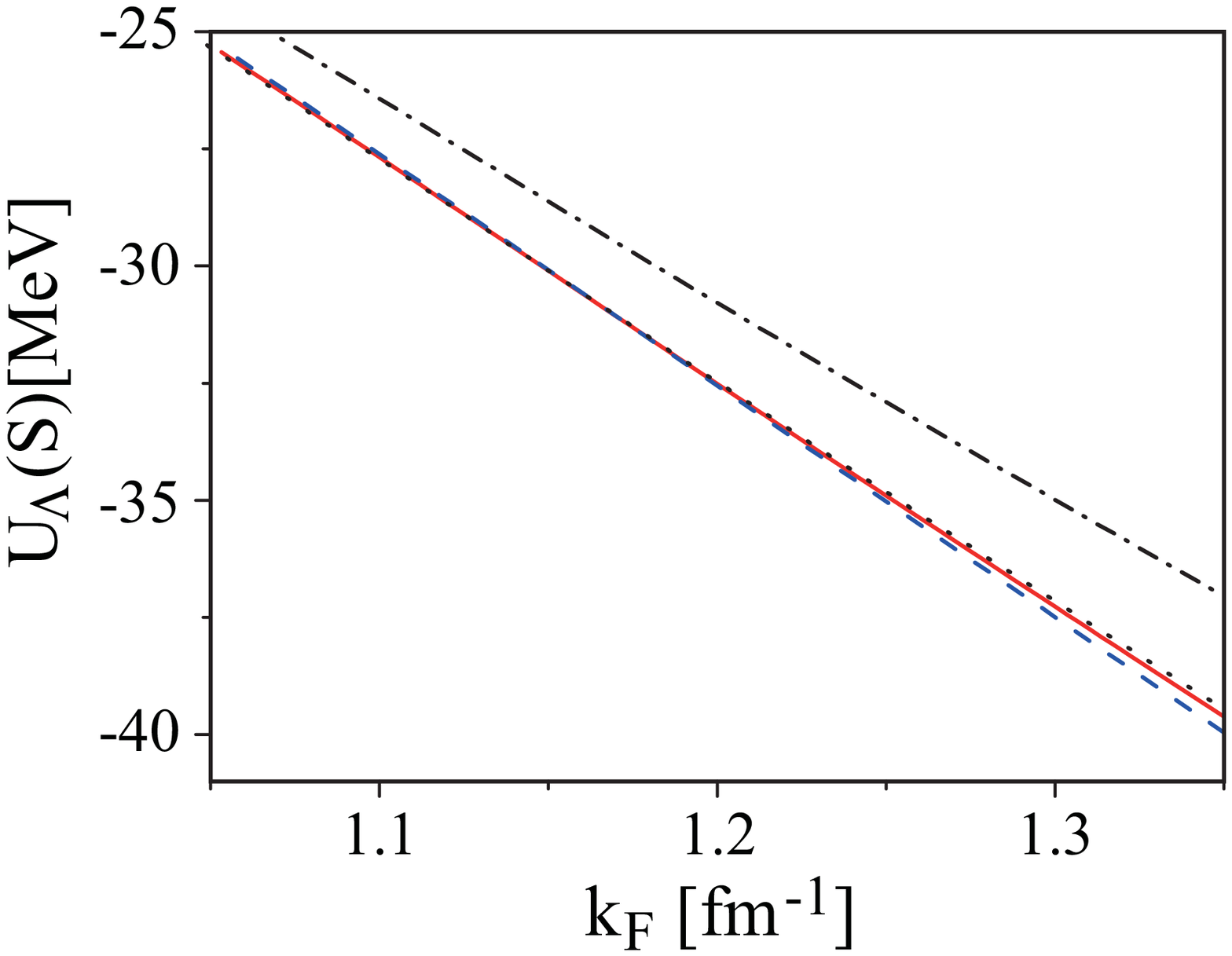} 
\caption{\small \label{ULs}
(Color online) $S$-state contributions to $U_\Lambda$ 
as a function of $k_F$.
Also see the caption of Fig.\ref{UL}.
}
\end{figure} 

\begin{figure}[htb] 
\centering 
\includegraphics[width=8cm]{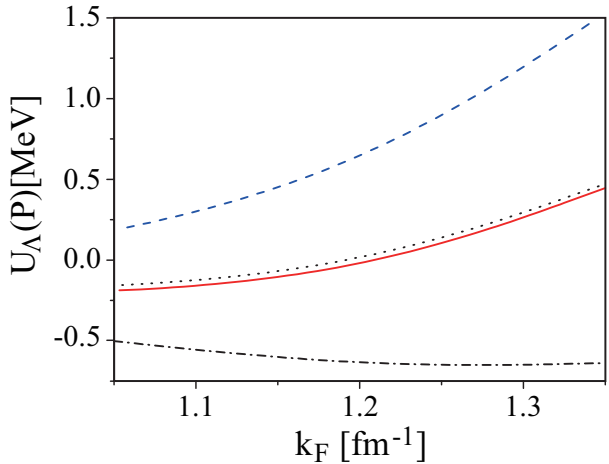} 
\caption{\small \label{ULp}
(Color online) $P$-state contributions to $U_\Lambda$ 
as a function of $k_F$.
Also see the caption of Fig.\ref{UL}.
}
\end{figure} 

In Fig.\ref{UL}, Fig.\ref{ULs} and Fig.\ref{ULp}, respectively,
$U_\Lambda$, $U_\Lambda(S)$ and $U_\Lambda(P)$ are drawn as a function
of $k_F$ in the cases of ESC08 models.  
Here, solid, dashed, dotted and dot-dashed curves are for
ESC14, ESC12, ESC08a and ESC08b.
It is found that the curves for ESC14 and ESC08a are very similar
to each other, and the main difference among those for 
ESC14, ESC08a, ESC12 and ESC08b is in the $P$-state contributions.
In Fig.\ref{ULp}, the important point is that the $P$-state contributions
for ESC14 and ESC08a are almost vanishing in the region of 
$k_F=1.1 \sim 1.2$ fm$^{-1}$. This feature appears also in case of ESC16,
and means that the $P$-state contributions are small for $B_\Lambda$ values
in light hypernuclei. On the other hand, there appears the repulsive
contribution substantially in the case of ESC12. 
As discussed later, the sizes of $P$-state contributions are related 
to a room to take MBE effects into account. 

For structure calculations of $\Lambda$ hypernuclei, we derive $k_F$-dependent 
effective local potentials ${\cal G}(k_F;r)$, simulating $\Lambda\!N$ $G$-matrices.
They are parameterized in a three-range Gaussian form: 
\begin{eqnarray}
{\cal G}(k_F,r)= \sum^3_{i=1}\, (a_i+b_i k_F +c_i k_F^2) \,
\exp {(-r^2/\beta_i^2)} \ .
\label{eq:GM7}
\end{eqnarray}
The parameters $(a_i, b_i, c_i)$ are determined so as to simulate 
the calculated $G$-matrix for each $^{2S+1}L_J$ state. 
The procedures to fit the parameters are given in Ref.~\cite{Yam10}.
The parameters for ${\cal G}(k_F,r)$ for ESC14 are given 
in Table~\ref{Gmat-2} .
It should be noted that ${\cal G}(k_F,r)$ are adjusted so as to reproduce 
exactly the values of $U_\Lambda$ in nuclear matter.
Contributions from $V_{MPP}^{(3)}$, $V_{MPP}^{(4)}$ and $V_{TBA}(r;\rho)$
are taken into account by modifying the second-range parts of ${\cal G}(k_F,r)$ 
by $\Delta {\cal G}(k_F,r)=(a+b k_F+c k_F^2) \exp -(r/\beta_2)^2$.

\begin{table}
\caption{Parameters in 
${\cal G}(k_F;r)= \sum_{i=1}^3 (a_i+b_i k_F+c_i k_F^2) \exp -(r/\beta_i)^2$
for ESC14. $a_i$ [MeV], $b_i$ [MeV$\cdot$fm], $c_i$ [MeV$\cdot$fm$^2$ ], and $\beta_i$ [fm] are given for each $i$.}
\label{Gmat-2}
\begin{ruledtabular}
\begin{tabular}{ccrrr}
&  $\beta_i$ & 0.50 &  0.90  &  2.00  \\
\hline
      & $a_i$  & $-$3434.  &    396.0  & $-$1.708  \\
$^1E$ & $b_i$  &    6937.  & $-$1057.  &   0.0    \\
      & $c_i$  & $-$2635.  &    415.9  &   0.0    \\
\hline
      & $a_i$  & $-$1933.  &    195.4  & $-$1.295  \\
$^3E$ & $b_i$  &    4698.  & $-$732.8  &   0.0    \\
      & $c_i$  & $-$1974.  &    330.1  &   0.0    \\
\hline
      & $a_i$  &    206.1  &    67.89  & $-$.8292  \\
$^1O$ & $b_i$  & $-$30.52  &    34.11  &   0.0    \\
      & $c_i$  &    16.23  &    2.471  &   0.0    \\
\hline
      & $a_i$  &    2327.  & $-$254.1  & $-$.9959  \\
$^3O$ & $b_i$  & $-$2361.  &    202.6  &   0.0    \\
      & $c_i$  &    854.3  & $-$43.71  &   0.0    \\
\end{tabular}
\end{ruledtabular}
\end{table}

\begin{table}
\caption{Parameters $a$ [MeV], $b$ [MeV$\cdot$fm], and $c$ [MeV$\cdot$fm$^2$] in
$\Delta {\cal G}(k_F;r)=  (a+b k_F+c k_F^2) \exp -(r/\beta_2)^2$
with $\beta_2=0.9$ fm
in the case of $g_P^{(3)}=2.34$, $g_P^{(4)}=30.0$ and $V_0=-21.0$.
}
\label{Gmat-3}
\vskip 0.2cm
\begin{ruledtabular}
\begin{tabular}{crrrr}
   &  $^1E$   & $^3E$    &  $^1O$   &  $^3O$  \\
\hline
 $a$ &    20.71 &    19.16 &    26.31 &    24.95  \\
 $b$ & $-$51.74 & $-$49.31 & $-$73.58 & $-$71.92  \\
 $c$ &    28.84 &    27.30 &    64.01 &    66.73  \\
\end{tabular}
\end{ruledtabular}
\end{table}

Here, $B_\Lambda$ values in finite systems are calculated by using
$\Lambda$-nucleus potentials in which $\Lambda\!N$ $G$-matrix ${\cal G}(k_F,r)$
interactions are folded into density distributions \cite{Yam10}.
Then, in order to treat $k_F$ values included in $G$-matrix interactions,
we use the averaged-density approximation (ADA), given as, 
\begin{eqnarray}
k_F =  \left( \frac{3\pi^2 \langle \rho \rangle}{2} \right)^{1/3},
\langle \rho \rangle = \int d^3r \rho_N(\textbf{r}) \rho_\Lambda(\textbf{r}),
\label{ADA}
\end{eqnarray}
where $\rho_N (\textbf{r})$ and $\rho_\Lambda (\textbf{r})$ represent 
the densities of the nucleons and $\Lambda$ particle, respectively.
In the next section, as well as in Ref. \cite{IYR16}, the HyperAMD is used
for structure calculations of $\Lambda$ hypernuclei based on ADA.
For spherical-core systems, it is confirmed that the present $G$-matrix
folding model and the HyperAMD give rise to similar results with each other.

Now, MPP and TBA parts are determined so that the experimental values 
of $B_\Lambda$ are reproduced by calculations with the
$G$-matrix folding model and the HyperAMD.
In the cases of ESC08a/b and ESC14, the experimental data can be 
reproduced well by varying only $V_{0}$ in TBA for values of  
$g_P^{(3)}=2.34$ and $g_P^{(4)}=30.0$ in the MPa set, being fixed 
to assure the stiffness of the neutron-star matter.
In Table~\ref{Gmat-3}, The parameters in $\Delta {\cal G}(k_F,r)$ are 
given in the case of $g_P^{(3)}=2.34$, $g_P^{(4)}=30.0$ and $V_0=-21.0$,
being adequate for ESC14.
The parameters in $\Delta {\cal G}(k_F,r)$ for ESC12 are given 
in Ref. \cite{IYR16}.

In the cases of ESC12 and NSC97e/f, it is needed to take
far smaller values of $g_P^{(3)}$ and $g_P^{(4)}$ for good fitting.
In the case of ESC04a, we obtain no reasonable set of
$g_P^{(3)}$, $g_P^{(4)}$ and $V_{0}$, which indicates that
the ESC04 models are inadequate to find reasonable MBE.
Table \ref{Gmat-4} gives determined values of
$g_P^{(3)}$, $g_P^{(4)}$ and $V_{0}$ and calculated values of
$\Delta B_\Lambda=
  B_\Lambda(^{89}_{\ \Lambda}$Y)$-B_\Lambda(^{16}_{\ \Lambda}$O)
  and $B_\Lambda(^{89}_{\ \Lambda}$Y) by the $G$-matrix folding model
for each interaction model.
Here, the values of $\Delta B_\Lambda$ are used to demonstrate 
roughly the mass dependence of $B_\Lambda$ values.
The values in parentheses are obtained without 
the MPP+TBA part $\Delta {\cal G}(k_F,r)$.
It should be noted that only in the case of ESC12 
the calculated values reproduces well the experimental values 
of $\Delta B_\Lambda$ and $B_\Lambda(^{89}_\Lambda$Y)
without contributions of $\Delta {\cal G}(k_F,r)$.

As found in Table~\ref{Gmat-1}, the order of $P$-state repulsions $U_\Lambda(P)$
is NSC97f$>$NSC97e$>$ESC12. This order corresponds to that of the attractions
$V_0$ in Table~\ref{Gmat-4}, where the stronger repulsion is needed to be cancelled by
the stronger attraction.

\begin{table}
\caption{$\Delta B_\Lambda$ [MeV] defined as $\Delta B_\Lambda=B_\Lambda(^{89}_\Lambda$Y)$-B_\Lambda(^{16}_\Lambda$O).
The experimental values of $\Delta B_\Lambda$ and
$B_\Lambda(^{89}_\Lambda$Y) are 10.7 MeV and 23.7 MeV, respectively.
Values in parentheses are calculated without MPP+TBA parts.
}
\label{Gmat-4}
\begin{ruledtabular}
\begin{tabular}{lccccc}
 & $g_P^{(3)}$  &  $g_P^{(4)}$ & $V_0$ & $\Delta B_\Lambda$ 
 & $B_\Lambda(^{89}_\Lambda$Y)  \\
  \hline
  ESC08a  & 2.34  & 30.0 & $-21.0$  & 10.9 (13.3) & 24.2 (26.6) \\
  ESC08b  & 2.34  & 30.0 & $-29.0$  & 10.9 (12.3) & 24.1 (24.2) \\
  ESC14   & 2.34  & 30.0 & $-21.0$  & 10.8 (13.2) & 24.0 (26.5) \\
  ESC12   & 0.39  &  0.0 &  $-5.0$  & 10.6 (10.8) & 23.9 (23.7) \\
  NSC97e  & 0.39  &  0.0 &  $-8.0$  & 10.4 (10.1) & 24.0 (22.8) \\
  NSC97f  & 0.0   &  0.0 & $-13.0$  & 10.3  (8.7) & 23.8 (20.2) \\
\end{tabular}
\end{ruledtabular}
\end{table}

\begin{table}
\caption{$\Delta B_\Lambda$ and $B_\Lambda(^{89}_\Lambda$Y) for ESC14 and ESC12.
Values in (a) are calculated with MPP+TBA, and values in
(b) and (c) are calculated only with MPP and TBA, respectively.
}
\label{Gmat-5}
\begin{ruledtabular}
\begin{tabular}{lcc}
 &  $\Delta B_\Lambda$  & $B_\Lambda(^{89}_\Lambda$Y)  \\
  \hline
  ESC14   &       &       \\
(a) MPP+TBA   & 10.8  & 24.0  \\
(b) MPP       &  7.9  & 17.9  \\
(c) TBA       & 16.1  & 33.4  \\
 
\hline
  ESC12   &       &       \\
(a) MPP+TBA   & 10.6  & 23.9  \\
(b) MPP       & 10.0  & 22.3  \\
(c) TBA       & 11.4  & 25.2  \\
\hline
\quad  exp     &  10.7 &  23.7 \\
\end{tabular}
\end{ruledtabular}
\end{table}

Table \ref{Gmat-5} gives calculated values of
$\Delta B_\Lambda$ and $B_\Lambda(^{89}_{\ \Lambda}$Y)
for ESC14 and ESC12.
Values in (a) are calculated with MPP+TBA, being the same
ones in the Table \ref{Gmat-4}.
Values in (b) and (c) are calculated only with MPP and TBA,
respectively.  In the case of ESC14, values of
$\Delta B_\Lambda$ and $B_\Lambda(^{89}_{\ \Lambda}$Y) 
including only MPP (TBA) are far smaller (larger) than
those including MPP+TBA. Thus, we find that the reasonable
values for MPP+TBA are owing to substantial canceling
between MPP and TBA contributions.
On the other hand, both contributions of MPP and TBA are small.

\begin{figure}[htb] 
\centering 
\includegraphics[width=8cm]{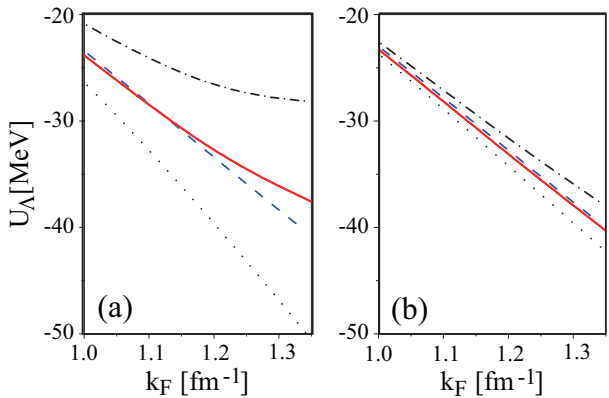} 
\caption{\small \label{Uden}
(Color online) (a) $U_\Lambda$ curves with ESC14. Solid (dashed) curve shows $U_\Lambda$ 
with (without) the MPP+TBA contributions. Dot-dashed (dotted) curve shows
$U_\Lambda$ only with MPP (TBA).
(b) Same as (a), but for ESC12. 
}
\end{figure}

From Figs.\ref{UL}, \ref{ULs}, and \ref{ULp},
some features can be found:
One is that the results for ESC14 and ESC08a are very similar
to each other. Their even-state parts give overbinding values
of $B_\Lambda$, where the odd-state parts are weak.  Then, MBE
plays a role to lift them up to experimental values. 
As shown in ref.\cite{ESC16}, this feature appears also 
in the case of ESC16.
On the other hand, in the case of ESC12 the even-state parts
are more attractive than those of ESC14 and ESC08a, and the strongly
repulsive odd-state parts contribute to reproduce the $B_\Lambda$ values,
and there remains only a small room for MBE to improve fitting.  
Then, it is very important that the odd-state contributions are
relatively smaller than the even-state ones in low-density region.
In the case of ESC12 the mass dependence of $B_\Lambda$ values, 
being estimated roughly by $\Delta B_\Lambda$, can be reproduced well 
owing to this feature of odd-state contributions.
In the left (right) side of Fig.\ref{Uden}, 
the $U_\Lambda$ curves are given for ESC14 (ESC12),
where solid (dashed) curves are with (without) the MPP+TBA
contributions. It is found that the contribution from the MPP+TBA part
in the ESC14 case is much larger than that in the ESC12 case.
The $k_F$ dependence of $U_\Lambda$ is (not) changed significantly 
by MBE in the former (latter) case.
Dot-dashed (dotted) curves in these figures show $U_\Lambda$ values 
only with MPP (TBA). In the case of ESC14, the solid curve with MPP+TBA 
turns out to be obtained by substantial canceling between MPP and TBA contributions.

As found in Fig.\ref{UL}, the $U_\Lambda$ curve for ESC08b is
similar to that for ESC12. However, the $U_\Lambda(S)$ values
for ESC08b are considerably less attractive than that for ESC12,
and $U_\Lambda(P)$ values for ESC08b (ESC12) are attractive 
(strongly repulsive). Due to this feature of ESC08b, the $B_\Lambda$ 
values are of rather underbinding, and the experimental values
are reproduced by adding the large attractive contributions 
from the MPP+TBA parts.
In the cases of NSC97e/f, the odd-state contributions are
more repulsive than those for ESC12 and there is 
no room of strong-MPP contribution.

Thus, it should be noted that a room for MBE is dependent sensitively
on the odd-state part in the $\Lambda\!N$ interaction, which has 
not yet been established experimentally in the present stage.

\section{Analysis of $B_\Lambda$ values based on HyperAMD}

In this Section, we discuss how the difference in $U_\Lambda$ of the $\Lambda\!N$ 
two-body interactions appears and affects the MBE in the systematics of $B_\Lambda$. 
As for the $V_{BB}$, we focus on ESC14 and ESC12 with MBE, because ESC08a/b are 
very similar to ESC14 as demonstrated in the previous section, and ESC12 is 
considerably different from ESC08a/b and ESC14 in odd states.
In this section, the calculations are performed with use of ESC14 and ESC12.
In our previous work \cite{IYR16}, based on ESC12, it was found that $B_\Lambda$ is sensitive to the structure of core nuclei, in particular core deformations. Furthermore, sophisticated treatment of $k_F$ related to core structure is also essential for quantitative discussion of $B_\Lambda$. 
In the present work, we perform structure calculations based on ESC14 within the framework of HyperAMD based on ADA from $^{9}_\Lambda$Li up to $^{59}_\Lambda$Fe, being compared with the results
in Ref. \cite{IYR16}.
The $G$-matrix interaction for ESC16 in Ref.\cite{ESC16} is considered to
give the result similar to that for ESC14.

\subsection{Framework of HyperAMD}

The Hamiltonian used in this study is 
\begin{align}
H = T_{N} + T_{\Lambda} - T_g + V_{NN} + V_{C} + V_{\Lambda N},
\end{align}
where $T_{N}, T_{\Lambda}$, and $T_{g}$ are the kinetic energies of the nucleons, $\Lambda$ particle, and center-of-mass motion, respectively. 
$V_{NN}$ and $V_{C}$ are the effective nucleon-nucleon ($NN$) and Coulomb interactions, respectively. The Coulomb interaction $V_{C}$ is approximated by the sum of seven Gaussians. As for the $\Lambda\!N$ interaction $V_{\Lambda N}$, we use the $G$-matrix interaction discussed above.

In this study we use the Gogny D1S force \cite{Gogny1,Gogny2} as the effective $NN$ interaction $V_{NN}$. 
In our previous work \cite{IYR16}, it was found that structure of the core nuclei affects the values of $B_\Lambda$. 
This fact tells us that proper description of core structure is indispensable to extract information of $\Lambda N$ interaction from the $B_\Lambda$ values in a model approach. 
Therefore, we need to use an appropriate effective $NN$ interaction in the HyperAMD calculation, which gives better agreement with the observed data in wide mass regions. 
The Gogny D1S force is one of such effective interactions.
It is found that the AMD calculation with Gogny D1S force successfully describes core deformations and gives reasonable binding energies of the core nuclei within a few percent of deviation from the observed data. 

The variational wave function of a single $\Lambda$ hypernucleus is described by the parity-projected wave function, $\Psi^\pm = \hat{P}^\pm \{ \mathcal{A} \{ \varphi_1,\ldots ,\varphi_A \} \otimes \varphi_\Lambda \}$, where
\begin{eqnarray}
\varphi_{i} \propto e^{ - \sum_{\sigma} \nu_\sigma \bigl(r_\sigma - Z_{i\sigma} \bigr)^2 } \otimes (u_i \chi_\uparrow + v_i \chi_\downarrow) \otimes (p \ {\rm or} \ n), \label{varphi}\\
\varphi_\Lambda \propto \sum_{m=1}^M c_m e^{- \sum_{\sigma} \nu_\sigma \bigl(r_\sigma - z_{m\sigma} \bigr)^2} \otimes (a_m \chi_\uparrow + b_m \chi_\downarrow).
\end{eqnarray}
Here the single-particle wave packet of a nucleon $\varphi_{i}$ is described by a single Gaussian, while that of $\Lambda$, $\varphi_\Lambda$, is represented by a superposition of Gaussian wave packets. 
The variational parameters $\bm{Z}_i$, $\bm{z}_m$, $\nu_\sigma$, $u_i$, $v_i$, $a_m$, $b_m$, and $c_m$ are determined to minimize the total energy under the constraint on the nuclear quadrupole deformation $(\beta, \gamma)$, and the optimized wave function $\Psi^\pm (\beta, \gamma)$ is obtained for each given $(\beta, \gamma)$.

\begin{figure}
  \begin{center}
    \includegraphics[keepaspectratio=true,width=86mm]{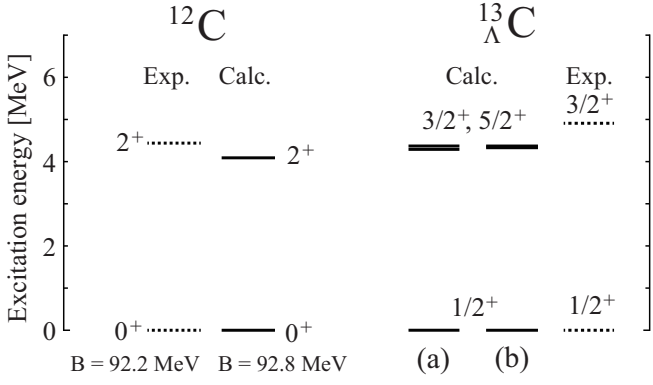}
  \end{center}
  \caption{Excitation spectra of $^{12}$C and $^{13}_\Lambda$C with (a) ESC12 + MBE and (b) ESC14 + MBE. Observed energy spectrum of $^{13}_\Lambda$C by the $\gamma$ ray spectroscopy experiments \cite{PRC65.034607(2002)} is also displayed.}
  \label{fig:Ex_C13L.eps}
\end{figure}

After the variation, we project out the eigenstate of the total angular momentum $J$ for each set of $(\beta,\gamma)$, 
\begin{align}
 \Psi^{J\pm}_{MK}(\beta,\gamma) &=
\frac{2J+1}{8\pi^2}\int d\Omega D^{J*}_{MK}(\Omega)R(\Omega) \Psi^\pm(\beta,\gamma).
\end{align}
The integrals over the three Euler angles $\Omega$ are performed numerically. 
Then the wave functions with different values of $K$ and $(\beta,\gamma)$ are superposed (generator coordinate method; GCM):
\begin{align}
 \Psi_n^{J\pm}&=\sum_p\sum_{K=-J}^{J} c_{nK} \Psi^{J\pm}_{MK}(\beta_p,\gamma_p),
 \label{eq:fullGCM}
\end{align}
where $n$ represents quantum numbers other than total angular momentum and parity. The coefficients $c_{npK}$ are determined by solving the Griffin-Hill-Wheeler equation. 
After the GCM calculation, we obtain both the ground and excited states of hypernuclei as shown in Fig. \ref{fig:Ex_C13L.eps}, where the present calculation nicely reproduces the observed spectra of $^{13}_\Lambda$C. 

In order to see the importance of describing the core deformation, we also perform the GCM calculation by using the intrinsic wave function with $\beta= 0.0$ only, namely,  
\begin{align}
 \Psi_n^{J\pm}&=\sum_{K=-J}^{J} c_{npK} \Psi^{J\pm}_{MK}(\beta=0.0), 
 \label{eq:spGCM}
\end{align}
and compare it with the usual GCM calculations given by Eq. (\ref{eq:fullGCM}), which was done in Ref. \cite{IYR16}. 

The $B_\Lambda$ is calculated as the energy difference between the ground states of a hypernucleus ($^{A+1}_{\Lambda}Z $) and the core nucleus $(^{A}Z)$ as $B_\Lambda = E(^{A}Z; j^\pm) - E(^{A+1}_{\Lambda}Z; J^\pm)$, where $E(^{A}Z; j^\pm)$ and $ E(^{A+1}_{\Lambda}Z; J^\pm)$ are calculated by GCM. 

We also calculate the squared overlap between the $\Psi^{J \pm}_{MK} ( \beta,\gamma)$ and GCM wave function $\Psi^{J \pm}_n$,
\begin{eqnarray}
O^{J\pm}_{MKn} ( \beta,\gamma ) = | \langle \Psi^{J \pm}_{MK} ( \beta,\gamma ) | \Psi^{J \pm}_n \rangle |^2,
\label{Overlap}
\end{eqnarray}
which is called the GCM overlap. 
$O^{J\pm}_{MKn} ( \beta,\gamma )$ shows the contribution of $\Psi^{J \pm}_{MK} ( \beta,\gamma )$ to each state $J^\pm$, which is useful to estimate the deformation of each state. In this study, we regard $(\beta, \gamma)$ corresponding to the maximum value of the GCM overlap as the nuclear deformation of each state.

\subsection{Impact of MBE on mass dependence of $B_\Lambda$}
\label{BLmdgs}

\begin{table*}
  \caption{
  Comparison of $-B_\Lambda$ [MeV] with including MBE by MPP + TBA based on ESC12 and ESC14. 
  Values of $B_\Lambda$ by using ESC12 with/without MBE are taken from Ref. I \cite{IYR16}. 
  $k_F$ [fm$^{-1}$] value calculated under ADA are also listed together with $\langle \rho \rangle$ [fm$^{-3}$] defined by Eq. (\ref{ADA}). 
  Observed values $B_\Lambda^{\rm exp}$ are taken from Refs. \cite{NPB52.1(1973),NPA83.306(1979),PRL66.2585(1991),NPA547.369(1992),NPA639.93c(1998),PRC64.044302(2001),NPA754.3(2005),PPNP57.564(2006),PRC90.034320(2014),Gogami}. 
  Values of $B^{\rm exp}_\Lambda$ with dagger are explained in text. 
  $\chi^2$ values calculated with $B_\Lambda$ and $B_\Lambda^{\rm exp}$ for the hypernuclei with $(\ast)$ are also listed. 
  The ground state spin-parity $J^\pi$ calculated and observed are also shown. 
  }
  \label{Tab:table1}
  \begin{ruledtabular}
  \begin{tabular}{ccccccccccccc}
 &         &          & \multicolumn{4}{c}{Based on ESC12 \cite{IYR16}}  &
                        \multicolumn{4}{c}{Based on ESC14} \\
 &         &          & \multicolumn{2}{c}{$V_{BB}$ only} & \multicolumn{2}{c}{w/ MBE} &
                        \multicolumn{2}{c}{$V_{BB}$ only} & \multicolumn{2}{c}{w/ MBE} & 
                        \multicolumn{2}{c}{Expt.} \\
 \cline{4-5}\cline{6-7} \cline{8-9}\cline{10-11} \cline{12-13} 
 & $\langle \rho \rangle$ & $k_F$ & $J^\pi$ & $-B_\Lambda$ & $J^\pi$ & $-B_\Lambda$ & $J^\pi$ &  $-B_\Lambda$ & $J^\pi$ & $-B_\Lambda$ & $J^\pi$ & $-B_\Lambda^{\rm exp}$ \\
 \hline
 $^{9}_\Lambda$Li$(\ast)$ & 0.072 & 1.02 & $5/2^+$ & $-7.9$ & $5/2^+$ & $-8.1$ & $5/2^+$ & $-7.6$ & $5/2^+$ & $-8.1$ & -- & $-8.50\pm0.12$\cite{NPA754.3(2005)}\\
 $^{9}_\Lambda$Be & 0.060 & 0.96 & $1/2^+$ & $-7.9$ & $1/2^+$ & $-8.1$ & $1/2^+$ & $-7.7$ & $1/2^+$ & $-8.1$ & $1/2^+$ & $-6.71\pm0.04$\cite{NPB52.1(1973)}\\
 $^{9}_\Lambda$B$(\ast)$  & 0.072 & 1.02 & $5/2^+$ & $-8.0$ & $5/2^+$ & $-8.2$ & $5/2^+$ & $-7.7$ & $5/2^+$ & $-8.2$ & -- & $-8.29\pm0.18$\cite{NPA754.3(2005)}\\
 $^{10}_\Lambda$Be$(\ast)$ & 0.077 & 1.04 & $2^-$ & $-8.7$ & $2^-$ & $-9.0$ & $2^-$ & $-8.6$ & $2^-$ & $-9.0$ & -- & $-9.11\pm0.22$\cite{NPA547.369(1992)},\\
  &  &  &  &  &  &  &  &  &  &  & &  $-8.55\pm0.18$\cite{Gogami}\\
 $^{10}_\Lambda$B$(\ast)$ & 0.075 & 1.04 & $2^-$ & $-8.9$ & $2^-$ & $-9.2$ & $2^-$ & $-8.7$ & $2^-$ & $-9.1$ & $1^-$\cite{NPA754.58c(2005),PRC41.1062(1990)} & $-8.89\pm0.12$\cite{NPB52.1(1973)}\\
 $^{11}_\Lambda$B$(\ast)$ & 0.081 & 1.05 & $7/2^+$ & $-9.8$ & $7/2^+$ & $-10.1$ & $7/2^+$ & $-9.7$ & $7/2^+$ & $-10.0$ & $5/2^+$\cite{NPA242.461(1975)} & $-10.24\pm0.05$\cite{NPB52.1(1973)}\\
 $^{12}_\Lambda$B$(\ast)$ & 0.083 & 1.07 & $2^-$ & $-11.0$ & $2^-$ & $-11.3$ & $2^-$ & $-11.0$ & $2^-$ & $-11.3$ & $1^-$\cite{NPA238.437(1975),NPA238.453(1975),NPA333.367(1980)} & $-11.37\pm0.06$\cite{NPB52.1(1973)},\\
  &  &  &  & & &  &  &  & &   &  & $-11.38\pm0.02$\cite{PRC90.034320(2014)}\\
 $^{12}_\Lambda$C$(\ast)$ & 0.086 & 1.08 & $2^-$ & $-10.7$ & $2^-$ & $-11.0$ & $2^-$ & $-10.8$ & $2^-$ & $-11.0$ & $1^-$\cite{NPA835.3(2010)} & $-10.76\pm0.19$\cite{NPA754.3(2005)}\\
 $^{13}_\Lambda$C$(\ast)$ & 0.090 & 1.10 & $1/2^+$ & $-11.3$ & $1/2^+$ & $-11.6$ & $1/2^+$ & $-11.5$ & $1/2^+$ & $-11.7$ & $1/2^+$ & $-11.69\pm0.19$\cite{NPA547.369(1992)}\\
 $^{14}_\Lambda$C$(\ast)$ & 0.093 & 1.11 & $0^-$ & $-12.4$ & $0^-$ & $-12.5$ & $0^-$ & $-12.4$ & $0^-$ & $-12.5$ & -- & $-12.17\pm0.33$\cite{NPA754.3(2005)}\\
 $^{15}_\Lambda$N & 0.098 & 1.13 & $1/2^+$ & $-12.6$ & $1/2^+$ & $-12.9$ & $1/2^+$ & $-12.9$ & $1/2^+$ & $-12.9$ & $3/2^+$\cite{NPA754.58c(2005)} & $-13.59\pm0.15$\cite{NPB52.1(1973)}\\
 $^{16}_\Lambda$O$(\ast)$ & 0.105 & 1.16 & $0^-$ & $-12.7$ & $0^-$ & $-13.0$ & $1^-$ & $-13.3$ & $1^-$ & $-13.0$ & $0^-$\cite{PRL93.232501(2004)} & $-12.96\pm0.05$\cite{NPA639.93c(1998)}$^\dag$\\
 $^{19}_\Lambda$O & 0.110 & 1.18 & $1/2^+$ & $-14.0$ & $1/2^+$ & $-14.3$ & $1/2^+$ & $-14.8$ & $1/2^+$ & $-14.3$ & -- & --\\
 $^{21}_\Lambda$Ne & 0.106 & 1.20 & $1/2^+$ & $-15.1$ & $1/2^+$ & $-15.4$ & $1/2^+$ & $-15.8$ & $1/2^+$ & $-15.5$ & -- & --\\
 $^{25}_\Lambda$Mg & 0.116 & 1.23 & $1/2^+$ & $-15.8$ & $1/2^+$ & $-16.1$ & $1/2^+$ & $-17.0$ & $1/2^+$ & $-16.1$ & -- & --\\
 $^{27}_\Lambda$Mg & 0.125 & 1.23 & $1/2^+$ & $-16.1$ & $1/2^+$ & $-16.3$ & $1/2^+$ & $-17.5$ & $1/2^+$ & $-16.2$ & -- & --\\
 $^{28}_\Lambda$Si & 0.125 & 1.23 & $2^+$ & $-16.4$ & $2^+$ & $-16.6$ & $2^+$ & $-17.8$ & $2^+$ & $-16.6$ & -- & $-17.1\pm0.02$\cite{PPNP57.564(2006),PRC53.1210(1996)}$^\dag$\\
 $^{32}_\Lambda$S$(\ast)$ & 0.130 & 1.24 & $0^+$ & $-17.4$ & $0^+$ & $-17.6$ & $1^+$ & $-19.4$ & $0^+$ & $-17.6$ & -- & $-18.0\pm0.5$\cite{NPA83.306(1979)}$^\dag$ \\
 $^{40}_\Lambda$K & 0.136 & 1.26 & $1^+$ & $-19.2$ & $1^+$ & $-19.4$ & $1^+$ & $-21.5$ & $1^+$ & $-19.4$ & -- & --\\
 $^{40}_\Lambda$Ca$(\ast)$ & 0.136 & 1.26 & $1^+$ & $-19.2$ & $1^+$ & $-19.4$ & $1^+$ & $-21.3$ & $1^+$ & $-19.3$ & -- & $-19.24\pm1.1$\cite{PRL66.2585(1991)}$^\dag$\\
 $^{41}_\Lambda$Ca & 0.136 & 1.26 & $1/2^+$ & $-19.4$ & $1/2^+$ & $-19.6$ & $1/2^+$ & $-21.5$ & $1/2^+$ & $-19.5$ & -- & --\\
 $^{48}_\Lambda$K & 0.141 & 1.27 & $1^+$ & $-20.1$ & $1^+$ & $-20.2$ & $1^+$ & $-22.6$ & $1^+$ & $-20.2$ & -- & --\\
 $^{51}_\Lambda$V$(\ast)$ & 0.151 & 1.31 & $11/2^+$ & $-20.4$ & $11/2^+$ & $-20.4$ & $11/2^+$ & $-23.5$ & $11/2^+$ & $-20.3$ & -- & $-20.51\pm0.13$\cite{PRC64.044302(2001)}$^\dag$\\
 $^{59}_\Lambda$Fe & 0.142 & 1.28 & $1/2^+$ & $-21.3$ & $1/2^+$ & $-21.4$ & $1/2^+$ & $-24.6$ & $1/2^+$ & $-21.7$ & -- & --\\
 \hline
 $\chi^2$ for $(\ast)$ &  &  & & 38.7 & & 3.61 & & 87.7 & &4.63 && \\
  \end{tabular}
  \end{ruledtabular}
\end{table*}

On the basis of ESC14 and ESC12, we discuss the effects of MBE on $B_\Lambda$ values. 
In Table \ref{Tab:table1}, the values of $B_\Lambda$ calculated with HyperAMD are summarized together with the experimental values of $B_\Lambda$ ($B_\Lambda^{\rm exp}$). 
It is noted that the values of $B_\Lambda^{\rm exp}$ with dagger are shifted deeper by 0.54 MeV from those reported in Refs. \cite{NPA83.306(1979),PRL66.2585(1991),NPA639.93c(1998),PRC64.044302(2001),PPNP57.564(2006)}, concerning the systematic difference of $B_\Lambda^{\rm exp}$ between the emulsion and $(\pi^+, K^+)$ (or $(K^-, \pi^-)$) experiments, which was pointed out by Ref. \cite{Gogami}. 
In Table \ref{Tab:table1}, we also show the $\chi^2$ values calculated by using the experimental and theoretical values of $B_\Lambda$ for the hypernuclei with asterisk to see the agreement of each other. 
As discussed in Ref. \cite{IYR16}, in cases of $^{9}_\Lambda$Be, $^{15}_\Lambda$N 
and $^{28}_\Lambda$Si, the calculated values of $B_\Lambda$ deviate considerably from the 
experimental values by the inherent reason for each system. 
Therefore, we exclude these hypernuclei from the evaluation of the $\chi^2$ values. 

Before the discussions on MBE, let us see the calculated values of $B_\Lambda$ 
without MBE ($V_{BB}$ only in Table \ref{Tab:table1}). 
Comparing ESC14 with ESC12, it is seen that the $B_\Lambda$ for ESC12 without MBE
are rather close to the experimental data. 
This is clearly seen in the $\chi^2$ values without MBE, 
i.e. the $\chi^2$ value for ESC12 ($\chi^2 = 38.7$) is much smaller than that 
for ESC14 ($\chi^2 = 87.7$).
However, this value for ESC12 is not extremely small, which indicates that there still exists 
a room to improve the fitting by adding MBE.
In the case of ESC14, the calculated values of $B_\Lambda$ are deviated much from the observations. 
In particular, in the medium-heavy hypernuclei, the $B_\Lambda$ values for ESC14 become larger than 
those with ESC12, and overestimate the $B_\Lambda^{\rm exp}$ considerably. 
For example, in $^{51}_\Lambda$V, the calculated value of $B_\Lambda$ for ESC14 is 23.5 MeV, 
whereas $B_\Lambda = 20.4$ MeV for ESC12 (\textit{cf.} $B_\Lambda^{\rm exp} = 20.51 \pm 0.13$ MeV \cite{PRC64.044302(2001)}). 
This is mainly due to the difference of the $P$-state interactions of ESC14 and ESC12.
In Fig. \ref{ULp} the $P$-state contribution for ESC14 is found to be far smaller than 
that of ESC12, while the $S$-state contributions for ESC14 and ESC12 are similar to each other.
Then, the difference of the $P$-state contributions for ESC14 and ESC12 appears more clearly 
in $B_\Lambda$ values of heavier hypernuclei than in those of lighter ones,
because $P$-state contributions are relatively small in light systems.

Next, let us discuss the MBE on the mass dependence of $B_\Lambda$. 
From the analysis in Sec. \ref{SecII}, we use different parameter sets of MBE combined with ESC14 and ESC12, as shown in Table \ref{Gmat-4}. 
In the case of ESC14, the MPP part of the MPa set ($g^{(3)}_{P} =2.34$ and $g^{(4)}_{P} = 30.0$) is used, which gives the stiff EoS of the neutron star matter. 
In ESC12 with MBE, the parameters $g^{(3)}_{P}$, $g^{(4)}_{P}$ and $V_0$ are determined so as to reproduce the $B_\Lambda^{\rm exp}$ in $^{16}_\Lambda$O 
without considering the stiffness of the EoS, 
and show nice agreement with $B_\Lambda^{\rm exp}$ in the wide mass regions \cite{IYR16}.
As a result, the strength of the MBE part combined with ESC12 is much weaker than that with ESC14. 
Hereafter, ESC14 (ESC12) combined with MBE is denoted as ESC14+MBE (ESC12+MBE). 

In Table \ref{Tab:table1}, the values of $B_\Lambda$ calculated with ESC14+MBE are also summarized together with those by using ESC12+MBE taken from Ref. \cite{IYR16}. 
It is found that the $B_\Lambda$ values for ESC14+MBE, as well as ESC12+MBE, reproduce 
the observed data within about 200 keV in the $9 \le A \le 59$ regions except for $^{9}_\Lambda$Be, 
$^{15}_\Lambda$N, and $^{28}_\Lambda$Si.
It is clearly seen that the $\chi^2$ values are reduced by including MBE ($\chi^2 = 4.63$ for ESC14+MBE and $\chi^2 = 3.61$ for ESC12+MBE), which means that the agreement of $B_\Lambda$ is improved significantly by including MBE. 
Here, it has no meaning to discuss the difference in the two small $\chi^2$ values, because we did 
not search these values exactly as minimum values for variation of the parameters included in MBE.

For the fine agreements of $B_\Lambda$, the MBE part plays an essential role, especially in ESC14+MBE, which is clearly seen in the comparison of $B_\Lambda$ between ESC14 and ESC14+MBE in Table \ref{Tab:table1}.
In the hypernuclei with $A \ge 16$, where the ESC14 causes overbinding of $B_\Lambda$, the MBE essentially acts as a repulsive force and shifts the $B_\Lambda$ to be close to the observed values. For example, in $^{51}_\Lambda$V, $B_\Lambda$ is shifted from 23.5 MeV to 20.3 MeV by adding MBE. 
In the light hypernuclei, the MBE gives attraction.
%
In ESC14+MBE, the MPP repulsion acts strongly at high density or large $k_F$, which gives the stiff EoS of the neutron-star matter, and becomes weaker as $k_F$ decreases, while TBA remains at small $k_F$ regions. Thus, the MBE gives attraction in the light hypernuclei with smaller values of $k_F$.

In the case of ESC12, the MBE brings about the minor changes of $B_\Lambda$ as seen in Table \ref{Tab:table1}. This is because the MPP and TBA combined with ESC12 are much weaker than those with ESC14. 
It is noted that the weak MPP in ESC12+MBE is inconsistent with the stiff EoS suggested by the massive neutron star. 
It is also found that if the strong MPP included in ESC14+MBE
is combined with ESC12, the derived values of $B_\Lambda$ contradict the observed data. 
Therefore, based on ESC12, there is no choice of MBE to satisfy both the observed data of $B_\Lambda$ and the stiff EoS of the neutron star matter. 
This indicates that the strong repulsion suggested by the massive neutron star imposes a stringent constraint on the $\Lambda\!N$ two-body interaction models. From the results in Table \ref{Tab:table1}, we conclude that the ESC14 is one of the $\Lambda\!N$ interaction models which satisfies these conditions. As discussed in the previous section, ESC08a/b are similar to ESC14 on this point.
$P$-state interactions in ESC14 and ESC08a/b are not strongly repulsive differently from ESC12,
and do not play a role to reproduce the mass dependence of $B_\Lambda$ values. Namely, there 
is a room for adding MBE with strong MPP repulsion owing to the weak $P$-state contribution.

\begin{figure}
  \begin{center}
    \includegraphics[keepaspectratio=true,width=86mm]{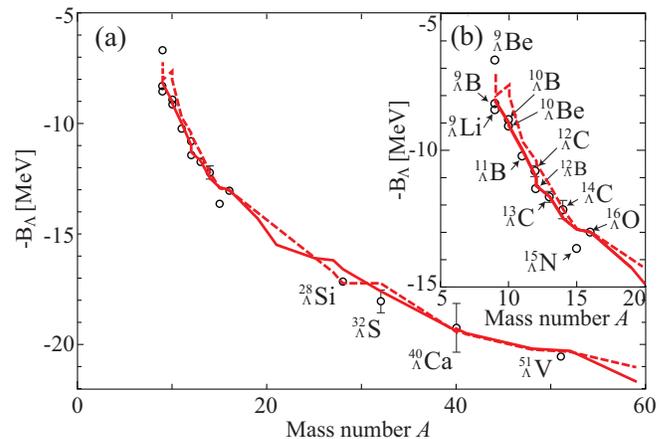}
  \end{center}
  \caption{(Color online) (a) Comparison of $B_\Lambda$ between full-basis GCM (solid) and spherical GCM (dashed) calculations. Open circles show observed data with mass numbers from $A=9$ up to $A=51$, which are taken from Refs. \cite{NPB52.1(1973),NPA83.306(1979),PRL66.2585(1991),NPA547.369(1992),NPA639.93c(1998),PRC64.044302(2001),NPA754.3(2005),PPNP57.564(2006),PRC90.034320(2014)}. 
$B_\Lambda^{\rm exp}$ measured by $(\pi^+, K^+)$ and $(K^-, \pi^-)$ reactions are shifted by 0.54 MeV as explained in text. (b) Same as (a), but magnified in the $5 \le A \le 20$ region. }
  \label{fig:fig1.eps}
\end{figure}

In the light hypernuclei, as pointed out in Ref. \cite{IYR16}, it is also important to describe properly the core structure, especially deformations of the core nuclei, to reproduce $B_\Lambda$, because it can affect the $B_\Lambda$ through the $k_F$ dependence of the interaction. 
In order to see the effects by core deformations, we compare the $B_\Lambda$ values calculated by performing the full-basis GCM (see Eq. (\ref{eq:fullGCM})) and spherical GCM (see Eq. (\ref{eq:spGCM})) calculations. 
In Fig. \ref{fig:fig1.eps}, it is seen that $B_\Lambda$ calculated by the spherical GCM are shallower than those in the full-basis GCM calculation and deviated from $B_\Lambda^{\rm exp}$, which is clearly seen in the $\chi^2$ values. In the spherical GCM calculation with ESC12+MBE, the $\chi^2$ is 171, whereas $\chi^2 = 4.63$ in the full-basis GCM calculation. 
This is because spherical states make $k_F$ larger through the increase of $\langle \rho \rangle$ in Eq. (\ref{ADA}). 
The difference of $B_\Lambda$ in Fig. \ref{fig:fig1.eps} is quite similar to the results with ESC12 in Ref. \cite{IYR16}. 

In Tab. \ref{Tab:table1}, the spin and parity $J^\pi$ are also listed for the ground states of the hypernuclei together with those known by the experiments. 
In case of the core nuclei having non-zero spin in the ground states, such as $^{11}$C and $^{10}$B, we naturally obtain the spin doublets in the corresponding hypernuclei, generated by coupling a $\Lambda$ particle with spin 1/2 to the ground states of the core nuclei. 
For example, in $^{12}_\Lambda$C, we obtain the $(1^-, 2^-)$ doublet corresponding to the $^{11}$C ground-state ($3/2^-$).
In Tab. \ref{Tab:table1}, it is seen that the calculated ground-state is $J^\pi = 2^-$, which is different from the observation ($J=1^-$). 
Similarly, in $^{11}_\Lambda$B, we obtain the $J^\pi = 7/2^+$ state as the ground state among the $(5/2^+, 7/2^+)$ doublet, whereas the $5/2^+$ state is the lowest in the experiment. 
This discrepancy is attributed to the property of the $\Lambda N$ spin-dependent interaction, which was discussed for the series of the ESC08 interaction models in Ref. \cite{Yam10}. 
In $^{12}_\Lambda$C, from the observed ground-state $1^-$, one can notice that the spin-singlet $\Lambda N$ interaction is slightly more attractive than the spin-triplet interaction. 
In the present calculation, it is considered that the spin-triplet part of the $\Lambda N$ interaction is slightly strong. 
Thus, the detailed properties of the spin dependence of the $\Lambda N$ interaction could affect the ordering of the ground-state doublet partners, though it has little influence on the $B_\Lambda$ curve. 

Finally, we also comment on the energy change of the nuclear part by the addition of a $\Lambda$ particle, namely the rearrangement energy $\Delta E_N$. 
Since $B_\Lambda$ is defined by the energy difference of the ground states between a hypernucleus and the core nucleus, it contains not only the $\Lambda$ single particle energy but also the energy changes of the core part, in which the former corresponds to the sum of the expectation values of $T_\Lambda$ and $V_{\Lambda\!N}$ and the latter is the $\Delta E_N$. Therefore, the rearrangement energy $\Delta E_N$ is defined as, 
\begin{align}
\label{eq:rearr}
\Delta E_N &= \frac{\langle \Psi^{J \pm}_n | H_N | \Psi^{J \pm}_n \rangle}{\langle \Psi^{J \pm}_n | \Psi^{J \pm}_n \rangle} - E(^AZ;j^\pi),\\
H_N &= T_{N} + V_{NN} + V_{C},
\end{align}
where $\Psi^{J \pm}_n$ is the GCM wave function defined by Eq. (\ref{eq:fullGCM}).
In Table \ref{tab:EN}, we summarize $\Delta E_N$ together with the ground-state energies of the core nuclei for the several hypernuclei with the different mass regions. It is found that the rearrangements energies are in order of a few hundred keV, which are quite smaller compared with $B_\Lambda$. Furthermore, it is also seen that $\Delta E_N$ is reduced as mass number increases. This is because adding a $\Lambda$ particle cannot change the core nuclei significantly, if the core nucleus is large enough. 

\begin{table}
\caption{
 Rearrangement energies $\Delta E_N$ [MeV], defined by Eq. (\ref{eq:rearr}), calculated for the ground states $J^\pi$ of the hypernuclei with different mass regions by using ESC14+MBE. Calculated ($E_{\rm core}$) and observed ($E^{\rm exp}_{\rm core}$) energies of the ground states $J^\pi_{\rm core}$in the core nuclei are also listed in MeV.
}
\label{tab:EN}
\begin{ruledtabular}
\begin{tabular}{cccccc}
  & $J^\pi$ & $\Delta E_N$ & $J^\pi_{\rm core}$ & $E_{\rm core}$ & $E_{\rm core}^{\rm exp}$\cite{CPC36.1287(2012)} \\
  \hline
  $^{9}_\Lambda$B   & $5/2^+$ & 0.24 & $2^+$ &  -41.5 &  -37.7 \\
  $^{13}_\Lambda$C  & $1/2^+$ & 0.16 & $0^+$ &  -92.8 &  -92.2 \\
  $^{28}_\Lambda$Si & $2^+$ & 0.14 & $5/2^+$ & -219.7 & -219.4 \\
  $^{41}_\Lambda$Ca & $1/2^+$ & 0.03 & $0^+$ & -341.7 & -342.0 \\
  $^{48}_\Lambda$K  & $1^+$ & 0.03 & $1/2^+$ & -400.2 & -389.0 \\
\end{tabular}
\end{ruledtabular}
\end{table}

\subsection{$B_\Lambda$ in $p$-states}
\label{Sec:BLmd-p}

\begin{table*}
\caption{Calculated (experimental) binding energy $B_\Lambda$ ($B_\Lambda^{\rm exp}$) and excitation energy $E_x$ ($E_x^{\rm exp}$) for the $p$-states $J^\pi$ in MeV together with $\rho$ [fm$^{-3}$] and $k_F$ [fm$^{-1}$] calculated by Eqs. (\ref{ADA}).  
$B_\Lambda^{\rm exp}$ with \dag\dag ($E_x^{\rm exp}$ with \dag\dag\dag) are calculated by using $E_x^{\rm exp}$ ($B_\Lambda^{\rm exp}$) of the $p$-states and the $B_\Lambda^{\rm exp}$ values shown in Table \ref{Tab:table1}. }
\label{tab:p-states}
\begin{ruledtabular}
\begin{tabular}{ccccccccccc}
  & & & \multicolumn{3}{c}{ESC12+MBE} & \multicolumn{3}{c}{ESC14+MBE} & \multicolumn{2}{c}{Exp.} \\
  \cline{4-6} \cline{7-9} \cline{10-11}
  & $\langle \rho \rangle$ & $k_F$ & $J^\pi$ & $-B_\Lambda$ & $E_x$ & $J^\pi$ & $-B_\Lambda$ & $E_x$ & $-B_\Lambda^{\rm exp}$ & $E_x^{\rm exp}$ \\
\hline
 $^{12}_\Lambda$B & 0.069 & 0.92 & $3^+$ & $-2.4$ & 8.8 & $3^+$ & $-2.4$ & 8.9 & $-1.289 \pm 0.048$ \cite{PRC90.034320(2014)} & $10.24 \pm 0.05$ \cite{PRC90.034320(2014)} \\
 $^{13}_\Lambda$C & 0.067 & 1.00 & $1/2^-$ & $-2.1$ & 9.5 & $1/2^-$ & $-2.4$ & 9.3 & $-0.9^{\dag\dag}$ & $10.830 \pm 0.087$ \cite{PRL86.4255(2001),PRC65.034607(2002)}\\
              &  &     &       &      &        &     &        &     & $-1.96^{\dag\dag}$ & $9.73 \pm 0.14$ \cite{NPA639.93c(1998)} \\
 $^{16}_\Lambda$O & 0.068 & 1.00 & $3^+$ & $-3.9$ & 9.1 & $3^+$ & $-4.0$ & 9.0 & $-2.39^{\dag\dag}$ & $10.57 \pm 0.06$ \cite{NPA639.93c(1998)} \\
 $^{28}_\Lambda$Si & 0.101 & 1.14 & $3^-$ & $-8.0$ & 8.6 & $3^-$ & $-8.4$ & 8.2 & $-7.5 \pm 0.2$ \cite{PRC53.1210(1996)}$^\dag$ & $9.6^{\dag\dag\dag}$ \\
 $^{51}_\Lambda$V & 0.124 & 1.22 & $11/2^-$ & $-12.7$ & 7.7 & $11/2^-$ & $-13.1$ & 7.2 & $-12.44 \pm 0.17$ \cite{PRC64.044302(2001)}$^\dag$ & 8.07$^{\dag\dag\dag}$ \\
\end{tabular}
\end{ruledtabular}
\end{table*}

\begin{table*}
\caption{Same as Table \ref{tab:p-states} but with a correction of $k_F$ as $k'_F = ( 1 + \alpha)k_F$. }
\label{tab:p-states2}
\begin{ruledtabular}
\begin{tabular}{cccccccccccc}
  && \multicolumn{4}{c}{ESC12+MBE} & \multicolumn{4}{c}{ESC14+MBE} & \multicolumn{2}{c}{Exp.} \\
  \cline{3-6} \cline{7-10} \cline{11-12}
  & $\langle \rho \rangle$ & $k'_F$ & $J^\pi$ & $-B_\Lambda$ & $E_x$ & $k'_F$ & $J^\pi$ & $-B_\Lambda$ & $E_x$ & $-B_\Lambda^{\rm exp}$ & $E_x^{\rm exp}$ \\
\hline
  && \multicolumn{4}{c}{$\alpha = 0.070$} & \multicolumn{4}{c}{$\alpha = 0.070$} & & \\
 $^{12}_\Lambda$B & 0.069 & 0.98 & $3^+$ & $-0.7$ & 10.6 & 0.98 & $3^+$ & $-0.8$ & 10.5 & $-1.289 \pm 0.048$ \cite{PRC90.034320(2014)} & $10.24 \pm 0.05$ \cite{PRC90.034320(2014)} \\
 $^{13}_\Lambda$C & 0.067 & 1.07 & $1/2^-$ & $-0.9$ & 10.8 & 1.07 & $1/2^-$ & $-1.0$ & 10.7 & $-0.9^{\dag\dag}$ & $10.830 \pm 0.087$ \cite{PRL86.4255(2001),PRC65.034607(2002)}\\
  &  &  &  &  & &  &  & & & $-1.96^{\dag\dag\dag}$ & $9.73 \pm 0.14$ \cite{NPA639.93c(1998)} \\
 $^{16}_\Lambda$O & 0.068 & 1.07 & $0^+$ & $-2.2$ & 10.8 & 1.07 & $2^+$ & $-2.6$ & 10.4 & $-2.39^{\dag\dag}$ & $10.57 \pm 0.06$ \cite{NPA639.93c(1998)} \\
\\ 
  && \multicolumn{4}{c}{$\alpha = 0.020$} & \multicolumn{4}{c}{$\alpha = 0.025$} & & \\
 $^{28}_\Lambda$Si & 0.101 & 1.17 & $3^-$ & $-7.4$ & 9.2 & 1.17 & $3^-$ & $-7.5$ & 9.1 & $-7.5 \pm 0.2$ \cite{PRC53.1210(1996)}$^{\dag}$ & 9.6$^{\dag\dag\dag}$ \\ 
\\
  && \multicolumn{4}{c}{$\alpha = 0.010$} & \multicolumn{4}{c}{$\alpha = 0.015$} & & \\
 $^{51}_\Lambda$V  & 0.124 & 1.24 & $11/2^-$ & $-12.3$ & 8.1 & 1.24 & $11/2^-$ & $-12.5$ & 7.8 & $-12.44 \pm 0.17$\cite{PRC64.044302(2001)}$^{\dag}$ & 8.07$^{\dag\dag\dag}$ \\
\end{tabular}
\end{ruledtabular}
\end{table*}

Let us focus on excited $p$-states of hypernuclei, 
in which the $\Lambda$ particle in $p$-orbit is bound to the ground state of the core nucleus. 
In general, since the $\Lambda$ particle in $p_{1/2}$ and $p_{3/2}$ orbits can couple to the core nuclei, several $p$ states appear. In this paper, we focus on the lowest $p$ state in excitation energy for each hypernucleus. 
In Sec. \ref{BLmdgs}, the HyperAMD calculation nicely reproduces the observed data of $B_\Lambda$ in the ground states of the hypernuclei using ESC12+MBE and ESC14+MBE by taking into account their structures, where the ADA treatment works well to obtain appropriate values of $k_F$ from the wave functions of the hypernuclei. In this section, we discuss the $B_\Lambda$ values in the $p$-states for the light ($^{12}_\Lambda$B, $^{13}_\Lambda$C, and $^{16}_\Lambda$O) and medium-heavy ($^{28}_\Lambda$Si and $^{51}_\Lambda$V) hypernuclei. 
In these hypernuclei, the $p$-states were observed in various experiments. The $(\pi^+, K^+)$ reaction experiments show the peak structure corresponding to the $p$-states in $^{13}_\Lambda$C \cite{NPA639.93c(1998)}, $^{16}_\Lambda$O \cite{NPA639.93c(1998)}, $^{28}_\Lambda$Si \cite{PRC53.1210(1996)}, and $^{51}_\Lambda$V \cite{PRC64.044302(2001)}. In $^{13}_\Lambda$C, the excitation energy of the $p$-states was precisely measured by the $\gamma$-ray spectroscopy experiment \cite{PRL86.4255(2001),PRC65.034607(2002)}. Recently, in $^{12}_\Lambda$B, the $(e,e'K^+)$ reaction experiment was performed at Thomas Jefferson National Accelerator Facility (JLab), which shows clear peaks regarded as the $p$-states with high resolution \cite{PRC90.034320(2014)}. 

Table \ref{tab:p-states} shows the calculated values of $B_\Lambda$ and the excitation energies 
$E_x$ of $p$-states with ESC12+MBE and ESC14+MBE together with the observed values.
The calculated values of $B_\Lambda$ are found to be slightly overbound by $0.3 \sim 1.1$ MeV
in comparison with the observed data. Now, let us try to modify the choice of $k_F$ values
in ADA so as to reproduce $B_\Lambda$ correctly, considering that the ADA might be changed
suitably for weak $\Lambda$ bound states:
We tune the $k_F$ value in ADA as $k'_F = (1 + \alpha)k_F$ by introducing a parameter $\alpha$, 
which is taken adequately for each mass region. For the light systems with $B_\Lambda \sim$ a few MeV, we take $\alpha = 0.070$ so as to reproduce the experimental values of $B_\Lambda$ in the $p$-state of $^{13}_\Lambda$C ($B_\Lambda^{\rm exp} = 0.9$ MeV). Here, the $B_\Lambda^{\rm exp}$ of $^{13}_\Lambda$C is obtained by subtracting the excitation energy $E_x^{\rm exp} = 10.830 \pm 0.087$ MeV measured by the $\gamma$-ray spectroscopy \cite{PRL86.4255(2001),PRC65.034607(2002)} from the $B_\Lambda^{\rm exp}$ in the ground state shown in Table \ref{Tab:table1}. 
The calculated values of $B_\Lambda$ and $E_x$ with $\alpha = 0.070$ are shown in Table \ref{tab:p-states2}. In the both cases with ESC12+MBE and ESC14+MBE, it is found that the $B_\Lambda^{\rm exp}$ and $E_x^{\rm exp}$ in $^{13}_\Lambda$C are reproduced using almost the same $\alpha$. 
In Table \ref{tab:p-states2}, it is seen that the values of $B_\Lambda$ and $E_x$ calculated with $\alpha = 0.070$ are much closer to the experimental values than those without $\alpha$ in $^{12}_\Lambda$B and $^{16}_\Lambda$O. 

In $^{28}_\Lambda$Si and $^{51}_\Lambda$V with stronger binding of $\Lambda$, it is found that smaller values of $\alpha$ give reasonable values of $B_\Lambda$ and $E_x$. Again, we tune $k_F$ so as to reproduce $B_\Lambda^{\rm exp}$ in the $p$-states of $^{28}_\Lambda$Si and determine $\alpha$ as $\alpha = 0.025$ ($\alpha = 0.020$) with ESC14+MBE (ESC12+MBE).
In $^{51}_\Lambda$V, it is also found that the $B_\Lambda^{\rm exp}$ and $E_x^{\rm exp}$ in the $p$-states are reproduced with $\alpha = 0.015$ ($\alpha = 0.010$) using ESC14+MBE (ESC12+MBE). These values of $\alpha$ in $^{28}_\Lambda$Si and $^{51}_\Lambda$V are much smaller than that determined in $^{13}_\Lambda$C.
Thus, larger values of $\alpha$ turn out to be needed, as $\Lambda$ bindings become weaker. It is worthwhile to point out, here, that the energy spectrum of $^{89}_{\ \Lambda}$Y can be reproduced nicely without the $\alpha$ parameter for the present $G$-matrix interactions with use of the $\Lambda$-nucleus folding model \cite{YFYR14}.

The degree of the modification of ADA by the $\alpha$ parameter is dependent on 
the smallness of $B_\Lambda$. 
As shown in Table \ref{tab:p-states2}, the $B^{\rm exp}_\Lambda$ values are in order of 1 MeV in $^{12}_\Lambda$B and $^{13}_\Lambda$C hypernuclei, which are much smaller than those in their ground states and the $p$-states of the medium-heavy hypernuclei. 
Therefore, $p$-state $\Lambda$ particles in light hypernuclei are rather weakly bound. 
Generally, in weakly bound states, since the $\Lambda$ particle distributes in broader region around the core nucleus, the $k_F$ value evaluated by Eq.(\ref{ADA}) under ADA could be smaller and then it makes the $B_\Lambda$ values larger. 
The above result shows that it brings about some overbinding of $\Lambda$ to use Eq.(\ref{ADA})
naively for weakly-bound $\Lambda$ states. Then, the $\alpha$ parameter plays a role to make $k_F$
larger and correct the overdoing of increase of $B_\Lambda$ by smaller $k_F$ values.
In heavier hypernuclei with increasing $B_\Lambda$, the above effect is less important, and thus $\alpha$ can be smaller. 
It is demonstrated in Table \ref{tab:p-states2} that there is a good correspondence between
the decrease of $B_\Lambda$ values and the increase of $\alpha$ values.
Thus, the present calculation reproduces the $B_\Lambda^{\rm exp}$ in $p$-states based on ADA with only a minor correction of $k_F$. 
This shows the validity of the HyperAMD calculations with the $G$-matrix interactions for applying to not only the ground states but also $p$-states of $\Lambda$ hypernuclei with large mass regions.

\subsection{$^{40}_\Lambda$K and $^{48}_\Lambda$K hypernuclei}

\begin{table*}
\caption{ Calculated values of $B_\Lambda$ [MeV] for the ground and $p$-states of $^{40}_\Lambda$K and $^{48}_\Lambda$K with ESC12 + MBE and ESC14 + MBE. In the $p$-states $k_F$ calculated by ADA is tuned as $k'_F = (1 + \alpha)k_F$ using $\alpha = 0.010$ and 0.020 ($\alpha = 0.015$ and 0.025) for ESC12 + MBE (ESC14 + MBE). $\langle \rho \rangle$ [fm$^{-3}$], $k'_F$ [fm$^{-1}$] and spin-parity $J^\pi$ are also listed.  }
\label{tab:K-hyper}
\begin{ruledtabular}
\begin{tabular}{ccccccccccc}
  & \multicolumn{5}{c}{Ground states} & \multicolumn{5}{c}{$p$-states} \\
  \cline{2-6} \cline{7-11}
  & $\langle \rho \rangle$ & $\alpha$ & $k'_F$ & $J^\pi$ & $-B_\Lambda$ & $\langle \rho \rangle$ & $\alpha$ & $k'_F$ & $J^\pi$ & $-B_\Lambda$ \\
\hline
 \multicolumn{9}{c}{ESC14 + MBE} \\
 $^{40}_\Lambda$K & 0.136 & 0.000 & 1.263 & $1^+$ & $-$19.4 & 0.109 & 0.015 & 1.191 & $2^-$ & $-$10.4 \\
                  &       &       &       &       &         & 0.109 & 0.025 & 1.202 & $2^-$ & $-$10.1 \\
 $^{48}_\Lambda$K & 0.141 & 0.000 & 1.278 & $1^+$ & $-$20.2 & 0.117 & 0.015 & 1.219 & $1^-$ & $-$11.6 \\
                  &       &       &       &       &         & 0.117 & 0.025 & 1.231 & $1^-$ & $-$11.3 \\
 \\
 \multicolumn{9}{c}{ESC12 + MBE} \\
 $^{40}_\Lambda$K & 0.136 & 0.000 & 1.263 & $1^+$ & $-$19.4 & 0.109 & 0.010 & 1.185 & $2^-$ & $-$10.2 \\
                  &       &       &       &       &         & 0.109 & 0.020 & 1.196 & $2^-$ & $-$9.9 \\
 $^{48}_\Lambda$K & 0.141 & 0.000 & 1.278 & $1^+$ & $-$20.2 & 0.117 & 0.010 & 1.213 & $1^-$ & $-$11.5 \\
                  &       &       &       &       &         & 0.117 & 0.020 & 1.225 & $1^-$ & $-$11.2 \\
\end{tabular}
\end{ruledtabular}
\end{table*}

At JLab, it is planned to perform the $(e,e'K)$ reaction experiment by using $^{40}$Ca and $^{48}$Ca as the target, i.e. potassium hypernuclei ($^{40}_\Lambda$K and $^{48}_\Lambda$K) are expected to be produced \cite{JLab}. 
As seen in Fig. \ref{fig:fig1.eps}, $B_\Lambda$ values were measured in several hypernuclei in this mass region. However, only a few observed data of $B_\Lambda$ are available. Furthermore, some of them have large ambiguities. For example, in $^{40}_\Lambda$Ca, the $B_\Lambda$ in the ground state has a large error (more than 1 MeV). 
Since absolute energies of hypernuclei can be measured with high resolution in the spectroscopy experiment at JLab, precise values of $B_\Lambda$ will be available for not only in the ground states but also excited states in $^{40}_\Lambda$K and $^{48}_\Lambda$K. 
Therefore, it is expected that the validity of the present calculation could be confirmed by comparing with the JLab experiments in heavier hypernuclei with $40 \le A < 50$. 

Recently, in this mass region, the effect by the isospin dependence of $\Lambda\!N\!N$ force is discussed \cite{PED}. By the auxiliary field diffusion Monte Carlo (AFDMC) calculation \cite{PED,JLab}, it is shown that if the isospin dependence exists, it affects the $B_\Lambda$ values in neutron-rich $\Lambda$ hypernuclei such as $^{48}_\Lambda$K due to the asymmetry of the proton and neutron numbers. 
In the present study, our many-body force is isospin-independent, which affects strongly 
on the mass dependence of $B_\Lambda$ even in neutron-rich hypernuclei. On the other hand,
in our $\Lambda\!N$ $G$-matrix interactions, the charge-dependent components included in the ESC model
are not taken into account. On the basis of our present modeling for the $\Lambda\!N$ interaction, 
we predict the values of $B_\Lambda$ in $^{40}_\Lambda$K and $^{48}_\Lambda$K in the ground and $p$-states.

In Table \ref{tab:K-hyper}, the calculated values of $B_\Lambda$ in the ground states with ESC12 + MBE and ESC14 + MBE are presented, which are taken from Table \ref{Tab:table1}, i.e. these values are calculated without modifying ADA. 
It is found that the $B_\Lambda$ value in $^{48}_\Lambda $K is larger than that in $^{40}_\Lambda$K. 
As seen in Fig. \ref{fig:fig1.eps}, these values are consistent with the mass dependence of $B_\Lambda$. 
Therefore, the $B_\Lambda$ obtained by the present calculation becomes larger as the mass number increases. 
In the calculation for the $p$-states of $^{40}_\Lambda$K and $^{48}_\Lambda$K, we introduce small parameter $\alpha$ as $k'_F = (1 + \alpha)k_F$ in the ADA treatment in the same manner as in Sec. \ref{Sec:BLmd-p}. 
From the results of $^{28}_\Lambda$Si and $^{51}_\Lambda$V, it is expected that the appropriate value of $\alpha$ is in between 0.015 and 0.025 (0.010 and 0.020) in the $p$-states of $^{40}_\Lambda$K and $^{48}_\Lambda$K with ESC14 + MBE (ESC12 + MBE). 
Therefore, we calculate the $B_\Lambda$ in the $p$-states using these values of $\alpha$. 
The resulting values of $B_\Lambda$ in the $p$-states are also summarized in Table \ref{tab:K-hyper}. 
In the case of ESC14 + MBE, it is found that the ambiguity of $B_\Lambda$ coming from the $\alpha$ parameter is only about 300 keV, and the $B_\Lambda$ values are predicted to be 10.1 MeV $\le B_\Lambda \le$ 10.4 MeV and 11.3 MeV $\le B_\Lambda \le$ 11.6 MeV for $^{40}_\Lambda$K and $^{48}_\Lambda$K, respectively. 
These values are in between those in $^{28}_\Lambda$Si and $^{51}_\Lambda$V, and are increased depending on the mass number. 
We find the same trend of $B_\Lambda$ with the ESC12 + MBE. 
These values of $B_\Lambda$ are expected to be compared with the future experiments at JLab, which could give us useful information on properties of hyperonic many-body force.

\section{Summary}

Basic quantities in hypernuclei are $\Lambda$ binding energies $B_\Lambda$
which lead to a potential depth $U_\Lambda$ in nuclear matter.
In spite of the longstanding development of studies for $\Lambda\!N$ interactions,
values of $U_\Lambda$ derived from various interaction models are substantially
different from each other: There still remain ambiguities of models due to
lack of (accurate) $Y\!N$ scattering data.

The stiff EoS giving the neutron-star mass of $2M_{\odot}$ suggests the existence 
of strongly repulsive many-body effect (MBE) in the high-density region. 
On the other hand, the hyperon mixing in neutron-star matter brings about
the remarkable softening of the EoS.
In order to solve this ``Hyperon puzzle", we consider that the repulsive MBE
works also in hyperonic channels.
As a specific model for MBE, the multi-pomeron exchange repulsion (MPP) 
is added to the two-body interaction together with the phenomenological three-body 
attraction (TBA).

We adjust MBE so as to reproduce the observed data of $B_\Lambda$. Then,
it is evident that the strength of MBE depends on the two-body interaction model.
Even among various versions of the Nijmegen interaction models (ESC08a/b, ESC12, ESC14, 
ESC04a, NSC97e/f), there are considerable differences with each other. 
Especially, important is the difference among the $P$-state contributions. 
In the cases of ESC14 and ESC08a/b, the $P$-state contributions are almost vanishing,
where the mass dependence of $B_\Lambda$ can be reproduced well by adding MBE with 
the strong MPP repulsion assuring the stiff EoS of hyperon-mixed neutron-star matter.
In the cases of ESC12 and NSC97e/f, the $P$-state contributions are 
substantially repulsive and helpful to reproduce the mass dependence of $B_\Lambda$:
There is no room to introduce the strong MPP repulsion consistently with 
the experimental data. In the case of ESC04a, the $P$-state contribution is strongly
attractive, and it is difficult to reproduce the mass dependence of $B_\Lambda$
by adding the present form of MBE.           

The $B_\Lambda$ values of hypernuclei with $9 \le A \le 59$ are analyzed
in the framework of HyperAMD with use of the $\Lambda\!N$ $G$-matrix interactions
derived from ESC14 and ESC12. In both cases, the calculated values of 
$B_\Lambda$ reproduce the experimental data within a few hundred keV, 
when MBE is taken into account. 
The values of $B_\Lambda$ and $E_x$ in $p$-states also can be reproduced well
by the HyperAMD, when the ADA is modified so as to make input values of
$k_F$ slightly larger for weakly-bound $\Lambda$ states. 
Though the results for ESC14 and ESC12 are quite
similar to each other, the strength of MPP repulsion included in MBE for ESC12
is far weaker than that for ESC14: The former (ESC14) is strong enough to give rise to
the stiff EoS of hyperon-mixed neutron star-matter, but the latter (ESC12) is not.

In the present, it is difficult to prove the existence of MBE including strong repulsion
on the basis of the experimental data of $B_\Lambda$, because the two-body interaction
model is not finely determined. However, we can say at least as follows:
The possible existence of the strong hyperonic repulsions suggested by the stiff EoS of
neutron stars is compatible with $\Lambda\!N$ interaction models giving
almost vanishing contributions of $P$-state interactions.

The Fortran codes ESC08c2012.f (ESC12), ESC08c2014.f (ESC14), and HNPOTESC16.f (ESC16) can be found on the permanent open-access website NN-Online: {\it http://nn-online.org}. 



\end{document}